\documentclass[fleqn,10pt]{wlscirep}
\usepackage[utf8]{inputenc}
\usepackage{adjustbox}
\usepackage[T1]{fontenc}
\usepackage{subcaption}
\usepackage{booktabs} 
\usepackage{siunitx} 
\usepackage{adjustbox}
\usepackage{cite}

\title{Self-supervised denoising of visual field data improves detection of glaucoma progression}

\author[1,*]{Sean Wu}
\author[3,*]{Jun Yu Chen}
\author[3]{Vahid Mohammadzadeh}
\author[3]{Sajad Besharati}
\author[5]{Jaewon Lee}
\author[3]{Kouros Nouri-Mahdavi}
\author[3*]{Joseph Caprioli}
\author[4*]{Zhe Fei}
\author[1,2,*]{Fabien Scalzo}

\affil[1]{Pepperdine University, Keck Data Science Insitute, Malibu, 90263, United States}
\affil[2]{University of California Los Angeles, Department of Computer Science, Los Angeles, 90024, United States}
\affil[3]{University of California Los Angeles, Stein Eye Institute, David Geffen School of Medicine, Los Angeles, 90095, United States}
\affil[4]{University of California Riverside, Department of Statistics, Riverside, 92521, United States}
\affil[5]{Columbia University, Department of Computer Science, New York, 10025, United States}

\affil[*]{fabien.scalzo@pepperdine.edu, zhe.fei@ucr.edu, caprioli@jsei.ucla.edu }

\keywords{Glaucoma Progression, Masked Autoencoders, Visual Field Tests}

\begin{abstract}
Perimetric measurements provide insight into a patient’s peripheral vision and day-to-day functioning and is the main outcome measure for identifying progression of visual damage from glaucoma. However, visual field data can be noisy, exhibiting high variance, especially with increasing damage. In this study, we demonstrate the utility of self-supervised deep learning in denoising visual field data from over 4000 patients to enhance its signal-to-noise ratio, and thus its ability to detect true glaucoma progression. We deployed both a variational autoencoder (VAE) and a masked autoencoder to determine which self-supervised model best smooths the visual field data while reconstructing salient features that are less noisy, and more predictive of worsening disease. Our results indicate that including a categorical p-value at every visual field location improves the smoothing of visual field data. Masked autoencoders led to cleaner denoised data than previous methods, such as variational autoencoders. A 4.7\% increase in detection of progressing eyes with pointwise linear regression (PLR) was observed. The masked and variational autoencoders’ smoothed data predicted glaucoma progression 2.3 months earlier when p-values were included compared to when they were not. The faster prediction of time to progression (TTP), and the higher percentage progression detected, support our hypothesis of masking out visual field elements during training while including p-values at each location would improve the task of detection of visual field progression. Our study has clinically relevant implications with regard to masking when training neural networks to denoise visual field data, resulting in earlier and more accurate detection of glaucoma progression. This denoising model can be integrated into future models for visual field analysis to enhance detection of glaucoma progression.

\end{abstract}

\begin{document}

\flushbottom
\maketitle

\section*{Introduction}
Glaucoma is a leading cause of irreversible blindness worldwide, encompassing a group of ocular disorders characterized by optic nerve damage, often associated with increased intraocular pressure (IOP), and without inappropriate intervention, can lead to slow progressive visual loss\cite{reis2012rates,heijl2009natural}. Automated perimetry is a standard method utilized utilized to measure visual field damage and serves as a pivotal diagnostic tool to track glaucoma's functional deterioration\cite{nouri2004predictive}. Efficient detection of the rates of change observed in perimetry is essential for glaucoma management, allowing for proper intervention and management adjustments for those most at highest risk\cite{jackson2023fast,besharati2024detecting}. 

Measurement of the rate of deterioration, with respect to a patient’s longevity, is used to guide various interventional strategies such as prescription eye drops,  laser treatment,  and various types of surgery, to reduce and maintain the intraocular pressure of the eye at a less-damaging level; therefore, monitoring glaucoma progression is of particular interest to ophthalmologists\cite{weinreb2014pathophysiology,lim2022surgical,stein2021glaucoma}. Standard automated perimetry (SAP) enables us to visualize key features of glaucoma's impact on the visual field, which are sometimes summarized by  by abnormalities on the outputtedsuch as pattern standard deviation (PSD) and total deviation (TD) plots. Although SAP has proven effective for for the detection of glaucoma and its progression, it provides noisy, imperfect data due to manual input from patients requiring  motor feedback, and many other factors \cite{camp2017will}. 

Prediction of glaucoma progression has become a deep-learning endeavor with structural measurements from optic disc photographs or optical coherence tomography (OCT) data \cite{moon2022deep,mohammadzadeh2023prediction,wu2023auxiliary, mohammadzadeh2024prediction}. Recently, machine learning of VF data have aided the detection of glaucoma progression; Shuldiner et al. utilized initial visual field exams to predict the risk of rapid glaucoma progression with deep feed-forward neural networks \cite{shuldiner2021predicting}. Another study leveraged multi-modal deep learning with visual field data and OCT data to predict final visual field data to predict the severity of progression. Other similar studies also deploy deep neural networks to predict final visual field data with a variety of methods and inputs. It is clear that visual field measurements are a cornerstone to better detect, understand, and predict progression and thus improve glaucoma care.

Longitudinal visual field data can exhibit high variability in terms of sensitivity. To address this issue, numerous studies have attempted to smooth visual field measurements\cite{rabiolo2019quantification,sabouri2022pointwise}, for example, with clustering and nearest neighbor averaging. Currently, two main approaches are used when smoothing or denoising visual field data. The first is the nearest neighbor smoothing, and the second is deploying deep neural networks, more specifically with variational autoencoders. Previously, research scientists have utilized classical unsupervised learning methods such as the variational Bayesian independent component analysis mixture model (vB-ICA-mm) to smoothen SAP data and further analyze visual field defects  by increasing signal-to-noise ratio.

In this study, we constructed an encoder-decoder masked-autoencoder to learn salient and useful features from visual field data to detect glaucoma progression. We compare our model with previous state-of-the-art deep neural networks, such as the variational autoencoder approach, and raw visual field data extracted directly from a clinical database. Furthermore, a categorical p-value at every visual field location is used to demonstrate an even more effective visual field denoising at the model output.

\begin{figure}[t]
\centering
\includegraphics[width=\linewidth]{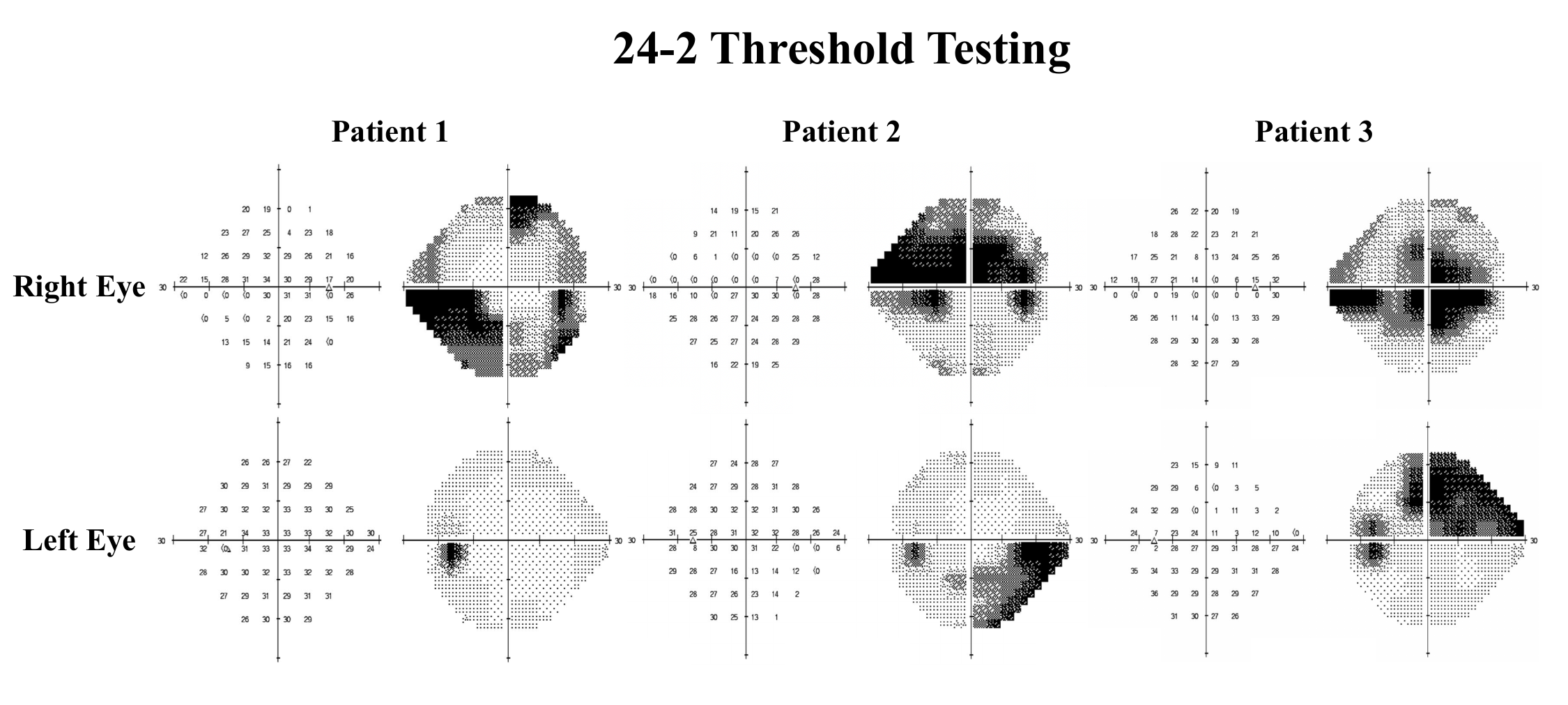}
\caption{Depiction of visual field exams from three different patients. The top row shows the right eye (OD) and the bottom row depics the left eye (OS), both depicting both the numerical and grayscale representations of central 24-2 threshold testing. }
\label{fig:VF}
\end{figure}

\section*{Methods}
\subsection*{Dataset Acquisition}
A total of 4,232 patients were included, with 16,924 visual field (VF) examinations. This type of data can be visualized in Figure 1. Each patient had a follow-up duration of at least 5 years, with a minimum of 6 visits. The baseline age had a mean (± SD) of 62.6 (± 13.6) years, and the mean follow-up period was 5.4 (± 4.6) years. This study was approved by UCLA’s Institutional Review Board (IRB) and followed the principles outlined in the Declaration of Helsinki as well as the policies established by the Health Insurance Portability and Accountability Act (HIPAA).  Visual field data were collected with a Humphrey Field Analyzer II (Carl Zeiss Meditec®, Dublin, CA). For each visual field exam, the reliability criteria were a false positive (FP) rate of $\leq$ 15\% and fixation loss (FL) and false negative (FN) rates of $\leq$ 30\%. Additionally, we exported the threshold sensitivity and total deviation (TD) values for each of the 54 visual field locations obtained from the 24-2 VF, and VFI was extracted as an XML file. Visual field locations 26 and 35 were omitted since these are blind spots for most eyes.

\begin{figure}[t]
\centering
\includegraphics[width=\linewidth]{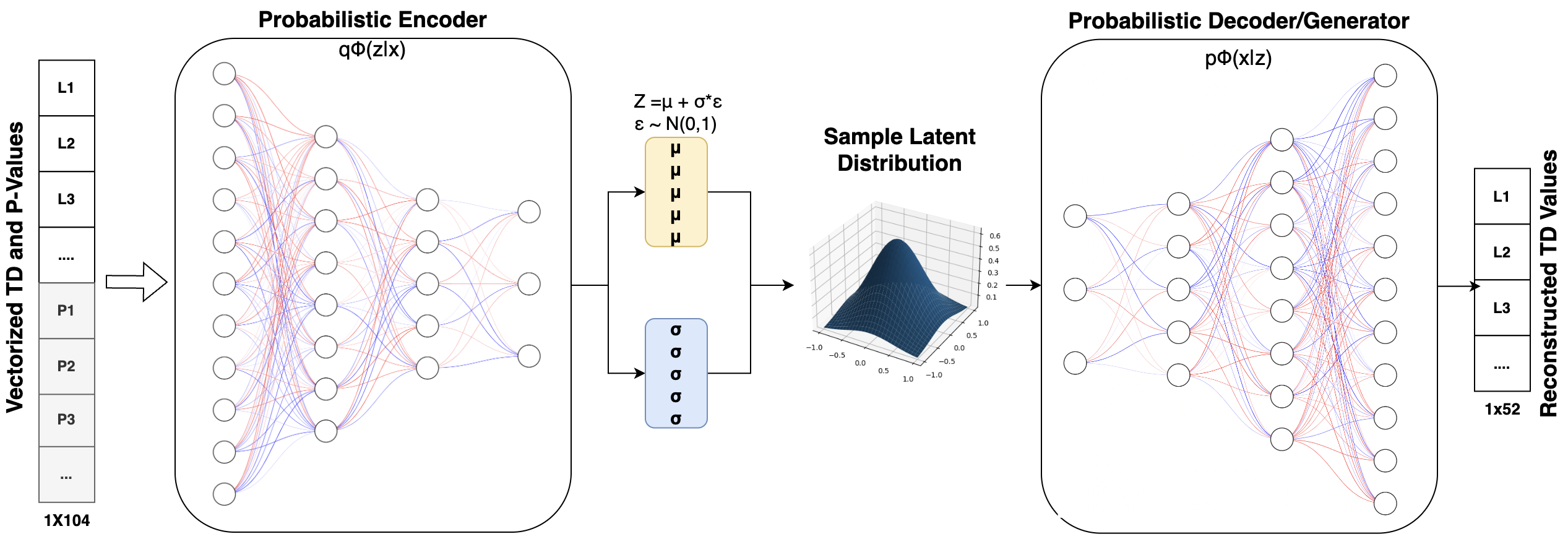}
\caption{Illustration of the VAE we deployed for this study, where we employed the basic probabilistic encoder-decoder architecture. Through the compression of the visual field data in the latent distribution, the goal is to make it difficult for the probabilistic decoder to reconstruct the noise in the data.}
\label{fig:VAE}
\end{figure}

\subsection*{Neural Denoising}
\begin{figure}[t]
\centering
\includegraphics[width=\linewidth]{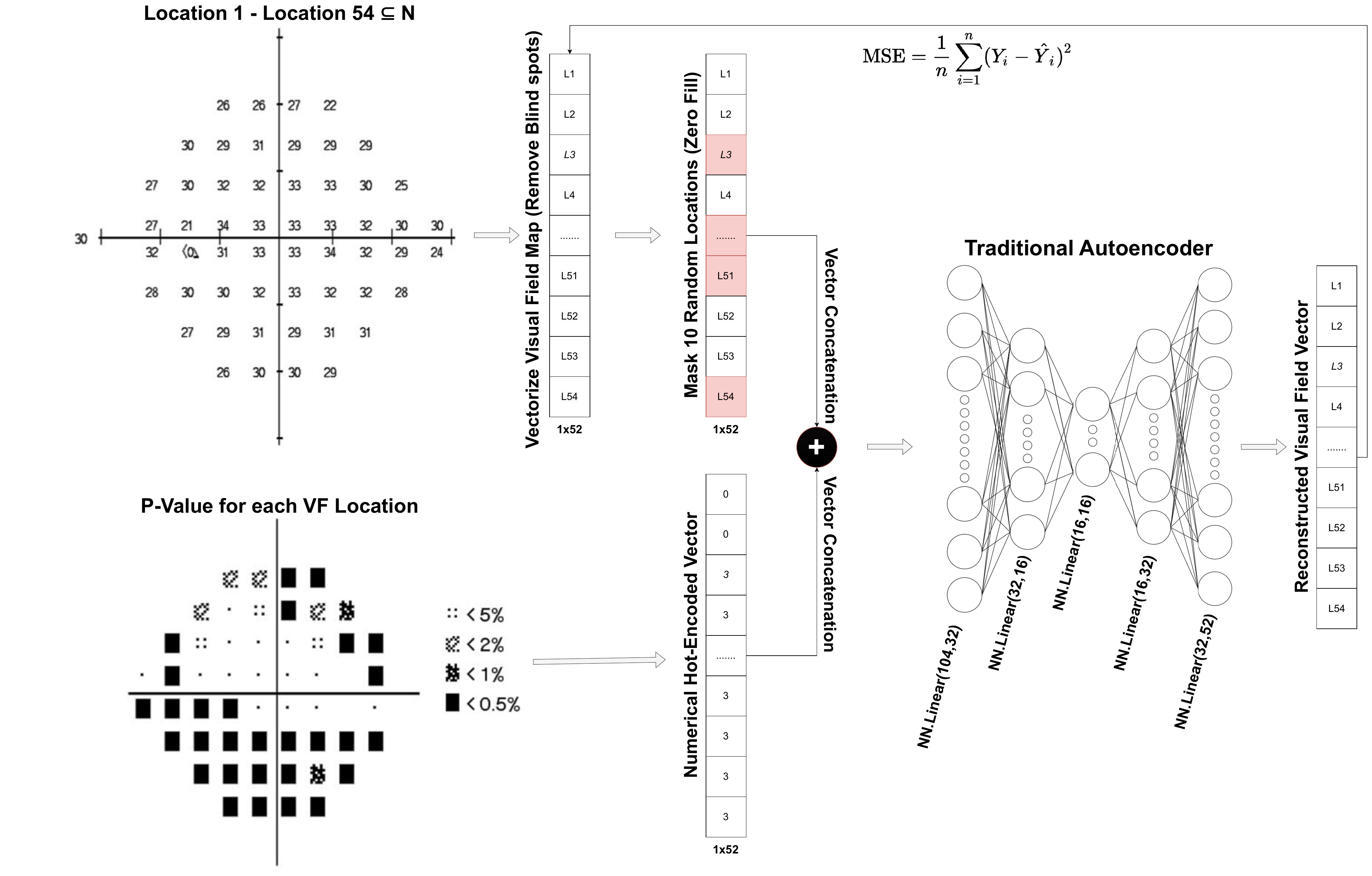}
\caption{Depiction of the masked autoencoder constructed. We first zero out 10 random visual field locations, then combine the vectorized visual field locations with the given p-values. Finally, we pass the data through a simple feed forward neural network to reconstruct the pre-masked data.}
\label{fig:MAE}
\end{figure}

We benchmarked two neural network architectures: the first one being the variational autoencoder and the second one being the masked autoencoder. The previous paper by Mohammadzadeh et al.\cite{mohammadzadeh2023efficacy} demonstrated the utility of deep learning, specifically VAEs, and their effectiveness in smoothing visual field data over traditional methods. In this paper, we replicate their results for VAE on our own dataset and demonstrate that the masked autoencoder architecture could be more useful in this case. For both models, we experimented with including a p-value at each location for improved denoising. Variational autoencoders (VAE) introduced by Kingma et. al\cite{kingma2013auto}, transformed the traditional autoencoder into a generative model by introducing a latent probability distribution parameterized a by mean \(\mu\) and standard deviation \(\sigma\). At a high level, the VAE encodes the input into a probability distribution, and the decoder or generator will sample from the latent space and reconstruct the data. Unlike a traditional autoencoder, VAEs are trained on the combined loss function of a typical reconstruction loss (Mean Squared Error, Mean Absolute Error etc), added to the Kullback Leibler divergence (KL-loss)\cite{zhang2021properties}. In our case, we chose a simple 5-layer VAE using a multi-layered perceptron or feed forward neural network. Out of the five layers, two layers were utilized for compressing the visual field and p-value data into a lower dimensional representation from 104 nodes to a hidden layer of 32 nodes and then 16 nodes, respectively, with a RelU nonlinear activation function between\cite{agarap2018deep}. We then have a bottleneck layer split into two separate outputs of 16 nodes of $\mu$ and 16 nodes of $\sigma$. The next step is to use the instantiated standard normal distribution N(0,1) to compute the sampled latent vector. We take the predicted $\mu$ vector and add it to the predicted $\sigma$ vector multiplied by the standard normal distribution shown in step 2 in the equations below. We then take the sampled vector $z$ and pass it to two dense layers of shape 16 to 32, 32 to 54 also with a RelU activation function as the upsampling layers.\\ 

\normalsize
\begin{align}
\text{Sample from $\mu$ and $\sigma$ of N:} \quad &z = \mu + \sigma \cdot N(\mu_{\text{shape}}) \\
\text{Compute Kullback-Leibler loss:} \quad &\text{KL Divergence} = \frac{1}{N} \sum_{i=1}^{N} \left( \sigma_i^2 + \mu_i^2 - \log(\sigma_i) - \frac{1}{2} \right)
\end{align}
\normalsize	
We computed the kullback leibler auxiliary loss by taking a summation from 1 to batch size of the predicted $\sigma$ squared added to the predicted $\mu$ squared subtracted by the logarithm of $\sigma$ - 1/2. The whole expression is divided by the batch size. The computation of this auxiliary loss is better visualized in step 3 of the equations above.

Another denoising autoencoder we constructed is a variant of the masked autoencoder first proposed by He et al\cite{he2022masked}. We created a simplified version with the idea of masking out parts of the input data to constrain the reconstruction process. Instead of using a vision transformer encoder (ViT)\cite{dosovitskiy2020image}, we employed a simple feedforward neural network as an encoder/decoder. The masked autoencoder pipeline is as follows (see Figure 4). For each training example, we first randomly "mask" or zero out ten random visual field locations from the original 52 numbers. Then, we concatenate this vector with 52 categorical p-values. We then pass the batch 1x104 vector through two encoding layers from 1x104 to 1x32 and 1x16, one bottleneck layer from 1x16 to 1x16, and finally, two decoding layers from 1x16 to 1x32 and then to 1x52. The loss is computed as the mean squared error between the predicted 52 numbers and the pre-masked 52 visual field numbers. Once the network is trained, during inference, we skip the masking step to "denoise" the raw input visual field data.

\subsection*{Training Details}
We trained both the VAE and masked autoencoder in a similar fashion to ensure a fair comparison between the two architectures. In this study, we trained a total of four neural networks: VAE, VAE + p-values, Masked autoencoder, and Masked autoencoder + p-values. We split our dataset into a standard training/validation/testing split and saved the epoch with the lowest validation loss to ensure proper generalization to unseen data and prevent overfitting. Some of the training parameters we used include a batch size of 32 to leverage parallel computation. We chose the Adam optimizer\cite{kingma2014adam} with a learning rate of 0.0001. To ensure a smooth and efficient training process, we deployed a learning rate scheduler algorithm that decreases the learning rate by a factor of 0.1 with a patience of 5 on the validation loss. Finally, we set a PyTorch manual seed of 42 to ensure each model has a similar weight initialization, and only one independent variable is changed at a time. Our code and data can be found in our open-source GitHub repository at the following link: \url{https://github.com/SeanWu25/Self-Supervised-Masked-Denoising-of-Visual-Field-Data}.

\subsection*{Progression Analysis Methods}
The VF data at each of the 52 test locations underwent noise reduction and reconstruction using both masked autoencoder and variational autoencoder (VAE) algorithms. We assessed glaucoma progression through multiple established methods: Mean Deviation (MD), Pointwise Linear Regression (PLR), and Glaucoma Rate Index (GRI). We compared the proportions of eyes demonstrating progression and time to progression with survival analysis across both denoised datasets (Masked Autoencoder and VAE) and raw data with and without p-value adjustments. Analyses were initiated from the baseline of 6 visits, progressively including additional visits up to the last follow-up. An eye was classified as progressing if the method’s criteria were met at two consecutive points and at the last follow-up.

We also performed survival analysis to examine the time until a specific event of interest. It is widely used in medical research for analyzing time-to-event data, where the event could be death, the occurrence of a disease, or, in our case, the progression of glaucoma. The concept of survival probability represents the likelihood of an individual or subject surviving beyond a certain time point without experiencing the specific event being studied. For the purposes of this paper, this directly translates to the probability of a patient not being diagnosed with worsening glaucoma. We also use the Kaplan-Meier curve to visually represent this data and illustrate the proportion of patients without glaucoma progression over a span of 20 years. Utilizing survival analysis provides invaluable insights into the effectiveness of our novel denoising technique on visual field data.

\subsubsection*{Pointwise Linear Regression (PLR)}

PLR is defined as the occurrence of a significant negative slope ($\leq -1$ dB/year, $P \leq 0.01$) in the regression of threshold sensitivity against time at any three of the 52 VF test locations.

\subsubsection*{Glaucoma Rate Index (GRI):}
The Glaucoma Rate Index (GRI) method, as introduced by Caprioli et al., evaluates each eye by allocating a score that reflects a spectrum ranging from significant deterioration to improvement. This process starts by applying pointwise exponential regression (PER) to calculate the pointwise rate of change (PRC), during which outliers are excluded, and adjustments are made for both age and location. In this framework, a model describing decay is represented as:
\begin{center}
    \large
    $S = e^{a + b \cdot FU}$
\end{center}
where b denotes the slope. Conversely, a model indicating improvement is formulated as:
\begin{center}
    \large
    $S_0 - S = e^{a + b \cdot FU}$
\end{center}
with \(S_0\) being the standard threshold value adjusted for age plus two standard deviations (SD), and \(b\) again representing the slope. PRC here is the yearly rate of change proportion, factoring in the entire range of the perimetric analysis and adjusted for both age and location variances. The GRI score is then computed by summing the PRC across all locations showing progression, and this sum is normalized to a scale ranging from -10 to +10. The analysis via the Pointwise Linear Regression (PLR) method entails regressing the sensitivity values (\(S\)) of each location against the follow-up period (\(FU\)) using simple linear regression (SLR). Visual field progression is thus determined based on the quantity of significantly deteriorating locations within the field.
\subsubsection*{Mean Deviation (MD):}
 MD measures the average deviation of the patient's sensitivity compared to age-matched normal values. It is a critical parameter where positive values indicate better-than-average vision, while negative values suggest impairment. MD typically ranges from +2 dB to -30 dB in reliable tests.

\section*{Simulations}

\subsection*{Simulation Setting}

To determine whether the proposed smoothing approaches would potentially address the noise without negatively affecting the true signal, we designed several  simulation VF data demonstrating different patterns of focal glaucoma progression and diffuse progression patterns, including age effect and cataract-induced sensitivity decrease. We performed the simulation algorithm previously published by our team\cite{rabiolo2019quantification}.

The R platform (R Foundation for Statistical Computing) was used to design the algorithm based on the models described previously.
The following steps were performed for this purpose:
\begin{itemize}
    \item The software defines the baseline threshold sensitivities, follow-up length, pointwise rates of progression annually, and the number of VF examinations. Tests are equally spaced over the follow-up period, and the ratio between the number of VFs and the length of the simulation was considered for defining the frequency of visits.

\item 
Linear regression was applied at each test location to calculate the VF values. An additional decay of 0.1 dB per year is added to simulate age-related decline, which was independent of the rate of progression.

\item Random noise was added to the simulated VF data in the same was as \cite{rabiolo2019quantification}.

\end{itemize}

The \texttt{visualFields} R package is able to generate the VFI, TD values, and the TD probability map for simulated data.
Five simulation settings, each with 376 eyes, including i) age-related decline, ii) slow progression, iii) medium progression, iv) fast progression, v) cataract rate were generated. Therefore, settings i) and iv) are non-glaucoma progressions and ii) - iv) are.
The simulation length was set at 9.5 years. Biannual testing frequency for a total of 20 VFs was defined. The baseline age was set at 60 years. Two VF examinations from actual glaucoma patients, one with a focal inferior nasal defect and another with a superior arcuate scotoma, were chosen as the 2 baselines. In the next step, 2 patterns of focal decay and diffuse decay were calculated as follows:

\begin{itemize}
    \item Focal decay, in which 4 (small scotoma), 8 (medium scotoma), or 16 (large scotoma) locations significantly deteriorate. Small defects were a nasal scotoma and a paracentral scotoma. Medium-sized defects consisted of a nasal step and an arcuate scotoma extending to 5 degrees from fixation. Large defects comprised of 2 broad inferior and superior arcuate scotomas. Three different rates of progression, $-0.5, -1.0$, and $-2.0$ dB per year, were applied to deteriorating locations in addition to normal age-related decay.

    \item Diffuse decay was defined as every location undergoing the same rate of progression.
\end{itemize}

Based on the baseline examinations, rates, and patterns of progression, 24 scenarios were simulated: eyes with focal decay, demonstrating deterioration of simulated scotomas, and eyes with diffuse decay were considered as progressing; eyes with simulated age-related and cataract-related decay were considered as nonprogressing.

\section*{Results}

The assessment of VF progression was conducted using five methods: raw noisy data, masked autoencoder (MAE) denoised data with p-values, masked autoencoder denoised data without p-values, variational autoencoder (VAE) denoised data with p-values, and variational autoencoder (VAE) denoised data without p-values. 

When evaluated using the Glaucoma Rate Index (GRI), the highest progression rates were observed in the masked autoencoder denoised data without p-values (77.13\%) and with p-values (68.95\%), followed by the VAE denoised data with p-values (62.38\%) and without p-values (59.31\%). In contrast, the raw noisy data demonstrated the lowest progression rate (50.60\%).

For the Mean Deviation (MD) metric, the raw data exhibited the highest progression rates across all categories, with nearly 100\% progression detection, likely due to noise effects amplifying the signal. The MAE denoised data showed consistently high progression rates with p-values (98.47\%) and without p-values (98.14\%), while the VAE denoised data showed lower progression rates, especially without p-values (72.61\%).

The Pointwise Linear Regression (PLR) method detected lower proportions of progressing eyes compared to MD and GRI. MAE denoised data without p-values showed the highest rates (67.49\%), followed closely by MAE with p-values (65.69\%). VAE denoised data exhibited the lowest progression rates under PLR, with and without p-values (36.90\% and 25.53\%, respectively). The raw data progression rate was moderate at 75.47\%.

These findings underscore the effectiveness of denoising techniques, particularly MAE, in enhancing the detection of VF progression while also reducing conversion time. The results highlight the importance of balancing denoising techniques to avoid overestimation or amplification of noise.


\begin{table}[ht]
\centering
\caption{Simulation results on 5 noisy VF data settings. Reported progression percentages by each method under 3 metrics: GRI, MD, and PLR. }
\begin{tabular}{lrrrrr}
  \hline
 &  Age-related decline & Slow progression & Medium progression & Fast progression & Cataract rate \\ 
  \hline
  \multicolumn{6}{c}{Progression by GRI} \\ 
Raw & 0.00 & 50.60 & 51.73 & 50.07 & 13.83 \\ 
MAE w/p & 2.39 & 68.95 & 69.68 & 68.35 & 43.09 \\ 
MAE & 0.27 & 77.13 & 76.00 & 75.33 & 59.57 \\ 
VAE w/p & 46.96 & 62.38 & 65.02 & 64.14 & 61.96 \\ 
VAE & 48.14 & 59.31 & 58.11 & 59.64 & 59.31 \\
  \hline
  \multicolumn{6}{c}{Progression by MD} \\ 
Raw & 1.60 & 99.93 & 99.93 & 99.93 & 100.00 \\ 
MAE w/p & 2.66 & 98.47 & 98.74 & 99.07 & 97.87 \\ 
MAE & 9.04 & 98.14 & 97.87 & 97.87 & 98.67 \\ 
VAE w/p & 9.57 & 65.09 & 67.35 & 68.02 & 39.10 \\ 
VAE & 9.84 & 72.61 & 69.48 & 70.68 & 42.29 \\ 
   \hline
  \multicolumn{6}{c}{Progression by PLR} \\ 
Raw & 0.00 & 75.47 & 73.01 & 73.20 & 0.00 \\ 
MAE w/p & 0.00 & 65.69 & 64.10 & 62.63 & 0.27 \\ 
MAE & 0.00 & 67.49 & 66.09 & 65.96 & 0.00 \\ 
VAE w/p & 0.00 & 36.90 & 36.77 & 34.44 & 0.53 \\ 
VAE & 0.80 & 25.53 & 26.60 & 26.33 & 3.99 \\ 
   \hline
\end{tabular}
\end{table}

\subsection*{Analysis of Kaplan-Meier Curves for Glaucoma Progression with Mean Deviation}
The Kaplan-Meier curve for Glaucoma Progression MD (Figure 2) presents a comparison of survival probabilities across five different dataset types over a period of 20 years. The survival probability is indicative of the proportion of subjects who have not been predicted with glaucoma progression at each time point. The initial survival probabilities are very similar across groups, with nearly all starting at 100\%. This uniformity underscores the absence of early progression or the efficacy of initial interventions. However, a divergence emerges as time elapses.  Notably, the autoencoder-derived methods demonstrate a decrement in survival probability, indicative of a higher glaucoma progression detection compared to the raw dataset. This phenomenon also aligns with the findings of our tables. Specifically, the masked autoencoder denoised datasets, both with and without p-value adjustments, exhibit a discernible reduction in survival probability relative to the variational autoencoder (VAE) counterparts. Despite this, the considerable overlap in confidence intervals suggests a potential lack of statistical significance when demarcating between the groups. After two decades of observation, the survival probabilities across all groups, with the exception of the raw data group, converges to a range between 50\% and 60\%. This convergence suggests that the distinction in the efficacy of various autoencoder methods for detecting glaucoma progression diminishes over an extended timeframe.

\subsection*{Analysis of Kaplan-Meier Curves for Glaucoma Progression with PLR}
\begin{figure}[t]
\centering
\includegraphics[width=\linewidth]{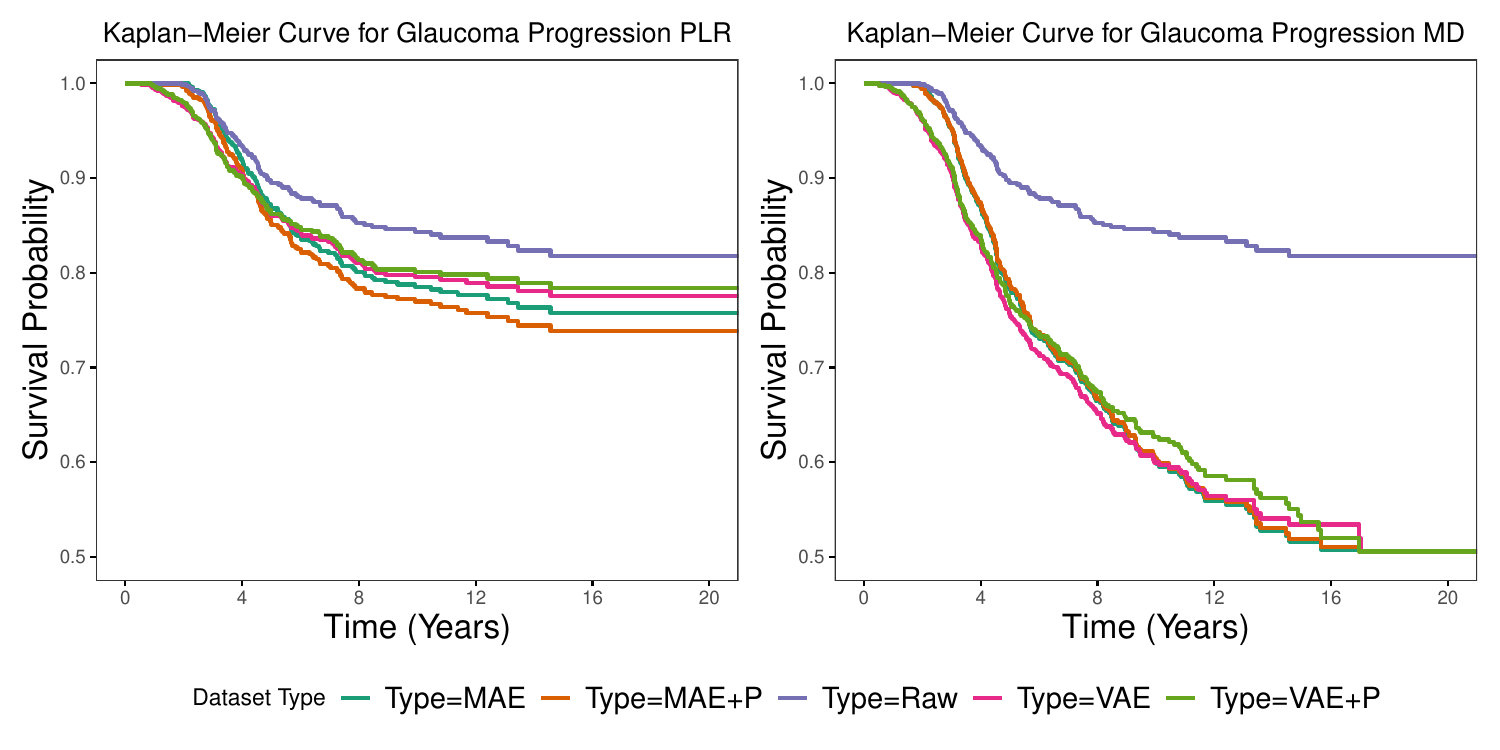}
\caption{Plot of Kaplan Meier survival curves for the mean deviation and pointwise linear regression metric showing that the masked-autoencoder with p-values outperform the VAE and raw data in survival analysis. Left and right are Kaplan Meier curves for PLR and MD respectively.}
\label{fig:KM}
\end{figure}

The Kaplan-Meier curve for Glaucoma Progression MD (Figure \ref{fig:KM}) presents a comparison of survival probabilities across five different dataset types over a period of 20 years. The survival probability is indicative of the proportion of subjects who have not been predicted with glaucoma progression at each time point. 
In the PLR dataset, a comparable pattern is seen, though the differences between the methods are less stark (Figure 2). After approximately five years, the survival probabilities begin to diverge, particularly between the raw data and the autoencoder-processed datasets. Around eight years and onwards, the masked autoencoder denoised datasets, with or without p-value adjustments, consistently show a lower survival probability, indicating a higher detection rate compared to their VAE counterparts. Among them, the masked autoencoder denoised dataset records the lowest survival probability. This pattern not only persists beyond the 20th year, highlighting a continuous difference in the effectiveness of detection methods. 

\begin{table}[htbp]
\centering
\caption{Visualization of glaucoma progression predicted by various models.}
\label{tab:glaucoma_progression}

\begin{subtable}{\textwidth}
\centering
\caption{Progression metrics for various models.}
\label{tab:progression_metrics}
\begin{tabular}{lccc}
\toprule
\textbf{Model / Progression} & \textbf{PLR (\%)} & \textbf{MD (\%)} & \textbf{GRI (\%)} \\
\midrule
Original VF Data & 14.48 & 36.67 & 15.21 \\
Masked Autoencoder & 19.62 & \textbf{37.08} & \textbf{17.05} \\
Masked Autoencoder + p-values & \textbf{21.21} & 36.51 & 14.50 \\
Variational Autoencoder & 16.54 & 33.41 & 6.00 \\
Variational Autoencoder + p-values & 17.08 & 32.00 & 7.93 \\
\bottomrule
\end{tabular}
\end{subtable}

\vspace{1em} 

\begin{subtable}{\textwidth}
\centering
\caption{Average conversion time to glaucoma progression for various models.}
\label{tab:avg_conv_time}
\begin{tabular}{lccc}
\toprule
\textbf{Model / Conv. Time} & \textbf{PLR (\%)} & \textbf{MD (\%)} & \textbf{GRI (\%)} \\
\midrule
Original VF Data & 7.80 & 7.89 & 6.72 \\
Masked Autoencoder & 7.20 & 7.80 & 7.47 \\
Masked Autoencoder + p-values & 7.01 & 7.80 & 7.38 \\
Variational Autoencoder & \textbf{6.07} & \textbf{7.21} & \textbf{4.50} \\
Variational Autoencoder + p-values & 6.26 & 7.32 & 5.39 \\
\bottomrule
\end{tabular}
\end{subtable}

\end{table}

\section*{Discussion and Conclusion}

We introduce two novel approaches to smoothing visual field data, a noisy measurement due in large part to the subjective nature of the test,  to more accurately detect glaucoma progression. Firstly, we demonstrate the effectiveness of masking several visual field locations during the network's training, providing a smoother data representation compared to standard glaucoma detection algorithms, such as PLR. This approach enables the detection of more glaucoma cases than pre-smoothing methods. Our method outperforms previous papers that utilize traditional variational autoencoders. Secondly, we incorporate a categorical p-value as a hot encoded vector in tandem with the vectorized visual field data. This addition provides supplementary input information to better encode the visual field data in a lower dimension. When comparing our masked autoencoder to the variational autoencoder, we observe a 4.3\% increase in the rate of progressing eyes detected by pointwise linear regression, a 4.51\% increase when using the mean deviation benchmark, and a 6.57\% increase when using the glaucoma rate index. The masked autoencoder outperforms the VAE with permutations in pointwise linear regression by 26.54\%, representing a significant increase in the number of progressing eyes detected by gold standard progression algorithms. We also demonstrate that including p-value input improves denoising, as evident in our best-masked autoencoder model, which predicts 1.59\% more progressing eyes and detects glaucoma progression 0.19 years faster with pointwise linear regression. Similarly, with the glaucoma rate index metric, including p-values helps the network produce smoothed data that predicts glaucoma progression 0.09 years faster than using raw data. These improvements from masked autoencoders + p-values are visualized in Table 1, depicting results for the PLR and mean deviation metrics.

While our research highlights the importance of the two denoising techniques for visual field data, there are areas where improvements can further be made. Incorporating longitudinal visual field data into a model may further enhance the quality of the time series. By utilizing recurrent neural networks, such as Long Short-Term Memories (LSTM) \cite{sak2014long}, we can leverage past and future visual field data to better smooth the current data based on learned patient input patterns. Additionally, to maximize the effectiveness of masked autoencoders in future studies, we propose incorporating the Transformer \cite{vaswani2017attention} encoder and decoder, as suggested in the original paper by He et al. Furthermore, potential areas for improvement include the utilization of Convolutional Neural Networks (CNN) \cite{o2015introduction} and deep learning \cite{lecun2015deep} on the grayscale representation of 24-2 central visual field tests (see Figure 1) for denoising. The image representation may contribute to a more effective diagnosis of glaucoma progression compared to the numerical format. This paper lays the groundwork for future research on predicting glaucoma progression using visual field data.

In conclusion, our research demonstrates that, prior to the application of techniques such as Pointwise Linear Regression (PLR) in clinical settings, passing visual field and categorical p-value data through our denoising models significantly reduces noise. This process yields a more effective data representation for cutting-edge glaucoma progression algorithms.

\bibliography{sample}

\begin{thebibliography}{10}
\urlstyle{rm}
\expandafter\ifx\csname url\endcsname\relax
  \def\url#1{\texttt{#1}}\fi
\expandafter\ifx\csname urlprefix\endcsname\relax\def\urlprefix{URL }\fi
\expandafter\ifx\csname doiprefix\endcsname\relax\def\doiprefix{DOI: }\fi
\providecommand{\bibinfo}[2]{#2}
\providecommand{\eprint}[2][]{\url{#2}}

\bibitem{reis2012rates}
\bibinfo{author}{Reis, A.~S.} \emph{et~al.}
\newblock \bibinfo{journal}{\bibinfo{title}{Rates of change in the visual field and optic disc in patients with distinct patterns of glaucomatous optic disc damage}}.
\newblock {\emph{\JournalTitle{Ophthalmology}}} \textbf{\bibinfo{volume}{119}}, \bibinfo{pages}{294--303} (\bibinfo{year}{2012}).

\bibitem{heijl2009natural}
\bibinfo{author}{Heijl, A.} \emph{et~al.}
\newblock \bibinfo{journal}{\bibinfo{title}{Natural history of open-angle glaucoma}}.
\newblock {\emph{\JournalTitle{Ophthalmology}}} \textbf{\bibinfo{volume}{116}}, \bibinfo{pages}{2271--2276} (\bibinfo{year}{2009}).

\bibitem{nouri2004predictive}
\bibinfo{author}{Nouri-Mahdavi, K.} \emph{et~al.}
\newblock \bibinfo{journal}{\bibinfo{title}{Predictive factors for glaucomatous visual field progression in the advanced glaucoma intervention study}}.
\newblock {\emph{\JournalTitle{Ophthalmology}}} \textbf{\bibinfo{volume}{111}}, \bibinfo{pages}{1627--1635} (\bibinfo{year}{2004}).

\bibitem{jackson2023fast}
\bibinfo{author}{Jackson, A.~B.} \emph{et~al.}
\newblock \bibinfo{journal}{\bibinfo{title}{Fast progressors in glaucoma: Prevalence based on global and central visual field loss}}.
\newblock {\emph{\JournalTitle{Ophthalmology}}} \textbf{\bibinfo{volume}{130}}, \bibinfo{pages}{462--468} (\bibinfo{year}{2023}).

\bibitem{besharati2024detecting}
\bibinfo{author}{Besharati, S.} \emph{et~al.}
\newblock \bibinfo{journal}{\bibinfo{title}{Detecting fast progressors: Comparing a bayesian longitudinal model to linear regression for detecting structural changes in glaucoma}}.
\newblock {\emph{\JournalTitle{American Journal of Ophthalmology}}} \textbf{\bibinfo{volume}{261}}, \bibinfo{pages}{85--94} (\bibinfo{year}{2024}).

\bibitem{weinreb2014pathophysiology}
\bibinfo{author}{Weinreb, R.~N.}, \bibinfo{author}{Aung, T.} \& \bibinfo{author}{Medeiros, F.~A.}
\newblock \bibinfo{journal}{\bibinfo{title}{The pathophysiology and treatment of glaucoma: a review}}.
\newblock {\emph{\JournalTitle{Jama}}} \textbf{\bibinfo{volume}{311}}, \bibinfo{pages}{1901--1911} (\bibinfo{year}{2014}).

\bibitem{lim2022surgical}
\bibinfo{author}{Lim, R.}
\newblock \bibinfo{journal}{\bibinfo{title}{The surgical management of glaucoma: A review}}.
\newblock {\emph{\JournalTitle{Clinical \& Experimental Ophthalmology}}} \textbf{\bibinfo{volume}{50}}, \bibinfo{pages}{213--231} (\bibinfo{year}{2022}).

\bibitem{stein2021glaucoma}
\bibinfo{author}{Stein, J.~D.}, \bibinfo{author}{Khawaja, A.~P.} \& \bibinfo{author}{Weizer, J.~S.}
\newblock \bibinfo{journal}{\bibinfo{title}{Glaucoma in adults—screening, diagnosis, and management: a review}}.
\newblock {\emph{\JournalTitle{Jama}}} \textbf{\bibinfo{volume}{325}}, \bibinfo{pages}{164--174} (\bibinfo{year}{2021}).

\bibitem{camp2017will}
\bibinfo{author}{Camp, A.~S.} \& \bibinfo{author}{Weinreb, R.~N.}
\newblock \bibinfo{journal}{\bibinfo{title}{Will perimetry be performed to monitor glaucoma in 2025?}}
\newblock {\emph{\JournalTitle{Ophthalmology}}} \textbf{\bibinfo{volume}{124}}, \bibinfo{pages}{S71--S75} (\bibinfo{year}{2017}).

\bibitem{moon2022deep}
\bibinfo{author}{Moon, S.}, \bibinfo{author}{Lee, J.~H.}, \bibinfo{author}{Choi, H.}, \bibinfo{author}{Lee, S.~Y.} \& \bibinfo{author}{Lee, J.}
\newblock \bibinfo{journal}{\bibinfo{title}{Deep learning approaches to predict 10-2 visual field from wide-field swept-source optical coherence tomography en face images in glaucoma}}.
\newblock {\emph{\JournalTitle{Scientific Reports}}} \textbf{\bibinfo{volume}{12}}, \bibinfo{pages}{21041} (\bibinfo{year}{2022}).

\bibitem{mohammadzadeh2023prediction}
\bibinfo{author}{Mohammadzadeh, V.} \emph{et~al.}
\newblock \bibinfo{journal}{\bibinfo{title}{Prediction of visual field progression with serial optic disc photographs using deep learning}}.
\newblock {\emph{\JournalTitle{British Journal of Ophthalmology}}}  (\bibinfo{year}{2023}).

\bibitem{wu2023auxiliary}
\bibinfo{author}{Wu, S.} \emph{et~al.}
\newblock \bibinfo{title}{Auxiliary-domain learning for a functional prediction of glaucoma progression}.
\newblock In \emph{\bibinfo{booktitle}{International Workshop on Ophthalmic Medical Image Analysis}}, \bibinfo{pages}{21--31} (\bibinfo{organization}{Springer}, \bibinfo{year}{2023}).

\bibitem{mohammadzadeh2024prediction}
\bibinfo{author}{Mohammadzadeh, V.} \emph{et~al.}
\newblock \bibinfo{journal}{\bibinfo{title}{Prediction of visual field progression with baseline and longitudinal structural measurements using deep learning}}.
\newblock {\emph{\JournalTitle{American Journal of Ophthalmology}}}  (\bibinfo{year}{2024}).

\bibitem{shuldiner2021predicting}
\bibinfo{author}{Shuldiner, S.~R.} \emph{et~al.}
\newblock \bibinfo{journal}{\bibinfo{title}{Predicting eyes at risk for rapid glaucoma progression based on an initial visual field test using machine learning}}.
\newblock {\emph{\JournalTitle{PloS one}}} \textbf{\bibinfo{volume}{16}}, \bibinfo{pages}{e0249856} (\bibinfo{year}{2021}).

\bibitem{rabiolo2019quantification}
\bibinfo{author}{Rabiolo, A.} \emph{et~al.}
\newblock \bibinfo{journal}{\bibinfo{title}{Quantification of visual field variability in glaucoma: implications for visual field prediction and modeling}}.
\newblock {\emph{\JournalTitle{Translational vision science \& technology}}} \textbf{\bibinfo{volume}{8}}, \bibinfo{pages}{25--25} (\bibinfo{year}{2019}).

\bibitem{sabouri2022pointwise}
\bibinfo{author}{Sabouri, S.}, \bibinfo{author}{Pourahmad, S.}, \bibinfo{author}{Vermeer, K.~A.}, \bibinfo{author}{Lemij, H.~G.} \& \bibinfo{author}{Yousefi, S.}
\newblock \bibinfo{journal}{\bibinfo{title}{Pointwise and region-wise course of visual field loss in patients with glaucoma}}.
\newblock {\emph{\JournalTitle{Translational Vision Science \& Technology}}} \textbf{\bibinfo{volume}{11}}, \bibinfo{pages}{20--20} (\bibinfo{year}{2022}).

\bibitem{mohammadzadeh2023efficacy}
\bibinfo{author}{Mohammadzadeh, V.} \emph{et~al.}
\newblock \bibinfo{journal}{\bibinfo{title}{Efficacy of smoothing algorithms to enhance detection of visual field progression in glaucoma}}.
\newblock {\emph{\JournalTitle{Ophthalmology Science}}} \bibinfo{pages}{100423} (\bibinfo{year}{2023}).

\bibitem{kingma2013auto}
\bibinfo{author}{Kingma, D.~P.} \& \bibinfo{author}{Welling, M.}
\newblock \bibinfo{journal}{\bibinfo{title}{Auto-encoding variational bayes}}.
\newblock {\emph{\JournalTitle{arXiv preprint arXiv:1312.6114}}}  (\bibinfo{year}{2013}).

\bibitem{zhang2021properties}
\bibinfo{author}{Zhang, Y.}, \bibinfo{author}{Liu, W.}, \bibinfo{author}{Chen, Z.}, \bibinfo{author}{Wang, J.} \& \bibinfo{author}{Li, K.}
\newblock \bibinfo{journal}{\bibinfo{title}{On the properties of kullback-leibler divergence between multivariate gaussian distributions}}.
\newblock {\emph{\JournalTitle{arXiv preprint arXiv:2102.05485}}}  (\bibinfo{year}{2021}).

\bibitem{agarap2018deep}
\bibinfo{author}{Agarap, A.~F.}
\newblock \bibinfo{journal}{\bibinfo{title}{Deep learning using rectified linear units (relu)}}.
\newblock {\emph{\JournalTitle{arXiv preprint arXiv:1803.08375}}}  (\bibinfo{year}{2018}).

\bibitem{he2022masked}
\bibinfo{author}{He, K.} \emph{et~al.}
\newblock \bibinfo{title}{Masked autoencoders are scalable vision learners}.
\newblock In \emph{\bibinfo{booktitle}{Proceedings of the IEEE/CVF conference on computer vision and pattern recognition}}, \bibinfo{pages}{16000--16009} (\bibinfo{year}{2022}).

\bibitem{dosovitskiy2020image}
\bibinfo{author}{Dosovitskiy, A.} \emph{et~al.}
\newblock \bibinfo{journal}{\bibinfo{title}{An image is worth 16x16 words: Transformers for image recognition at scale}}.
\newblock {\emph{\JournalTitle{arXiv preprint arXiv:2010.11929}}}  (\bibinfo{year}{2020}).

\bibitem{kingma2014adam}
\bibinfo{author}{Kingma, D.~P.} \& \bibinfo{author}{Ba, J.}
\newblock \bibinfo{journal}{\bibinfo{title}{Adam: A method for stochastic optimization}}.
\newblock {\emph{\JournalTitle{arXiv preprint arXiv:1412.6980}}}  (\bibinfo{year}{2014}).

\bibitem{sak2014long}
\bibinfo{author}{Sak, H.}, \bibinfo{author}{Senior, A.~W.} \& \bibinfo{author}{Beaufays, F.}
\newblock \bibinfo{journal}{\bibinfo{title}{Long short-term memory recurrent neural network architectures for large scale acoustic modeling}}.
\newblock {\emph{\JournalTitle{Google Interspeech 2014}}}  (\bibinfo{year}{2014}).

\bibitem{vaswani2017attention}
\bibinfo{author}{Vaswani, A.} \emph{et~al.}
\newblock \bibinfo{journal}{\bibinfo{title}{Attention is all you need}}.
\newblock {\emph{\JournalTitle{Advances in neural information processing systems}}} \textbf{\bibinfo{volume}{30}} (\bibinfo{year}{2017}).

\bibitem{o2015introduction}
\bibinfo{author}{O'Shea, K.} \& \bibinfo{author}{Nash, R.}
\newblock \bibinfo{journal}{\bibinfo{title}{An introduction to convolutional neural networks}}.
\newblock {\emph{\JournalTitle{arXiv preprint arXiv:1511.08458}}}  (\bibinfo{year}{2015}).

\bibitem{lecun2015deep}
\bibinfo{author}{LeCun, Y.}, \bibinfo{author}{Bengio, Y.} \& \bibinfo{author}{Hinton, G.}
\newblock \bibinfo{journal}{\bibinfo{title}{Deep learning}}.
\newblock {\emph{\JournalTitle{nature}}} \textbf{\bibinfo{volume}{521}}, \bibinfo{pages}{436--444} (\bibinfo{year}{2015}).

\end{thebibliography}

\section*{Acknowledgements}
We would like to thank the Keck Foundation for their grant to Pepperdine University to support our Data Science program and this research.

\section*{Author contributions statement}

F.S., J.C., and K.N.M. conceived the experiment(s),  S.W. conducted the experiment(s), V.M contributed on glaucoma and machine learning, J.Y.C. and Z.F. analysed the results. J.L helped with statistical analysis. S.W, V.M, J.C, J.L, Z.F, and F.S all helped create the manuscript. All authors reviewed the manuscript.

\end{document}